\documentclass[aps,prl,preprint]{revtex4}

\usepackage{epsfig}

\draft

\begin{document}
\author{Hong-shi Zong$^{1,2}$\footnote{E-mail:
zonghs@chenwang.nju.edu.cn}, Jia-lun Ping$^{3}$\footnote{E-mail:
jlping@njnu.edu.cn}, Wei-min Sun$^{1}$\footnote{E-mail:
sunwm@chenwang.nju.edu.cn},
Fang Wang$^{1}$\footnote{E-mail:
fgwang@chenwang.nju.edu.cn}, and 
Chao-hsi Chang$^{2,4,1}$\footnote{E-mail: zhangzx@itp.ac.cn}}
\address {$^{1}$ Department of Physics, Nanjing University, Nanjing 210093, P. R. China}
\address{$^{2}$ CCAST(World Laboratory), P.O. Box 8730, Beijing 100080, P. R. China}
\address{$^{3}$ Department of physics, Nanjing Normal University, Nanjing 210097, P. R. China}
\address{$^{4}$ Institute of Theoretical Physics, Academia Sinica, P.O. Box 2735, Beijing 100080, P. R. China}

\title{The Chiral Quark Condensate and the Bag Constant with Finite Chemical Potential}

\begin{abstract}
A method for obtaining the low chemical potential dependence of the dressed quark propagator from an effective quark-quark interaction model is developed. From this the chemical potential dependence of the chiral quark condensate and bag constant are evaluated. It is found that the chiral order parameter decreases with the increasing chemical potential $\mu$, which is qualitatively different from the results given by some authors in the literature. The reason for this discrepancy is discussed. 

\bigskip


\bigskip


\end{abstract}

\maketitle

High-energy heavy-ion collision experiments, which produce systems with large baryon density, are an important step in the search for quark-gluon plasma[1]. Present efforts are concentrated on exploring the domain of nonzero baryon density. The field theoretical study of finite baryon density in QCD begins with the inclusion of a chemical potential, $\mu$, which modifies the fermion piece of the Euclidean action: $\gamma\cdot\partial+m\rightarrow\gamma\cdot\partial-\gamma_{4}\mu+m$. The introduction of a chemical potential leads to a domain 
where the nonperturbative information can only be obtained from models(The numerical simulations of lattice-QCD action are not tractable in the domain of $\mu\not= 0$[2]). Since the global color symmetry model(GCM)[3,4,5] provides a nonperturbative framework that admits the simultaneous study of dynamical chiral symmetry breaking and confinement, it is expected to be well suited to explore the transition from hadronic matter to QGP[6].

It is the aim of this letter to study the chemical potential dependence of the dressed-quark propagator in the framework of the GCM, which provides a means of determining the behavior of the chiral and deconfinement order parameters. Up to this end let us consider the Euclidean action of the GCM at finite chemical potential $\mu$(In the case of the chiral limit):
\begin{eqnarray}
&&S_{GCM}[\bar{q},q;\mu]\nonumber\\
&&=\int d^{4} x \left\{\bar{q}(x)[\gamma\cdot\partial_{x}-\mu\gamma_{4}]q(x)\right\}+\int d^{4}x d^{4} y\left[\frac{g^2_{s}}{2} j^{a}_{\mu}(x) D^{ab}_{\mu\nu}(x-y)j^{b}_{\nu}(y)\right],
\end{eqnarray}
where $j^{a}_{\mu}(x)=\bar{q}(x)\gamma_{\mu}\frac{\lambda^{a}_{c}}{2}q(x)$  denotes the color octet vector current and $g^2_{s}D^{ab}_{\mu\nu}(x-y)$ is the dressed model gluon propagator in GCM. We want to stress here that $g^2_{s}D^{ab}_{\mu\nu}(x-y)$ does not evolve with $\mu$. For convenience, we will employ a model ansatz $D^{ab}_{\mu\nu}(x-y)=\delta_{\mu\nu}\delta^{ab}D(x-y)$ for the gluon propagator, which is often referred to as the so-called ``Feynman-like'' gauge[4](It should be noted that the above ansatz should be regarded merely as a model form for the gluon two-point function).

Introducing an auxiliary bilocal field $B^{\theta}(x,y)$ and applying the standard bosonization procedure the partition function of GCM[3,4]
\begin{equation}
{\cal{Z}}[\mu]=\int{\cal{D}} \bar{q}{\cal{D}}q e^{-S_{GCM}[\bar{q},q;\mu]}
\end{equation}
can be rewritten in terms of the bilocal fields $B^{\theta}(x,y)$
\begin{equation}
{\cal{Z}}[\mu]=\int{\cal{D}} B^{\theta} e^{-S_{eff}[B^{\theta};\mu]}
\end{equation}
with the effective bosonic action
\begin{equation}
S_{eff}[B^{\theta};\mu]=-Tr ln {\cal{G}}^{-1}[B^{\theta};\mu]+\int d^{4}x d^{4}y\frac{B^{\theta}(x,y) B^{\theta}(y,x)}{2 g^{2}_{s} D(x-y)}
\end{equation}
and the quark operator
\begin{equation}
{\cal{G}}^{-1}[B^{\theta};\mu]=[\gamma\cdot\partial_{x}-\mu\gamma_{4}]\delta(x-y)+\Lambda^{\theta} B^{\theta}(x,y).
\end{equation}
The matrices $\Lambda^{\theta}=D^{a}\otimes C^{b}\otimes F^{c}$ is determined by Fierz transformation in Dirac, color and flavor space of the current current interaction in Eq.(1), and are given by
\begin{equation}
\Lambda^{\theta}=\frac{1}{2}\left\{1_{D},i\gamma_{5},\frac{i}{\sqrt[]{2}}\gamma_{\nu},\frac{i}{\sqrt[]{2}}\gamma_{\nu}\gamma_{5}\right\}\otimes\left\{\frac{4}{3} 1_{C},\frac{i}{\sqrt[]{3}}\lambda^{a}_{C}\right\}\otimes\left\{\frac{1}{\sqrt[]{3}}1_{F},\frac{1}{\sqrt[]{2}}\lambda^{a}_{F}\right\}.
\end{equation}
One might expect that the complete set of the 16 Dirac matrices $\{1_{D}, \gamma_{5}, \gamma_{\mu}, \gamma_{5}\gamma_{\mu}, \sigma_{\mu\nu}\}$ must be employed in the description. However, by limiting the gluon two-point function $g^2_{s}D^{ab}_{\mu\nu}(x-y)$ to diagonal components in Lorentz indices, the tensor $\sigma_{\mu\nu}$ is excluded.

In the mean-field approximation, the fields $B^{\theta}(x,y)$ are substituted simply by their vacuum value $B^{\theta}_{0}(x,y)$, which is defined as
$\frac{\delta{S}_{eff}}{\delta{B}}{\mid}_{B_{0}}=0$ and is given by
\begin{equation}
B^{\theta}_{0}[\mu](x,y)=g^{2}_{s}D(x-y)tr[\Lambda^{\theta}
{\cal{G}}_{0}[\mu](x,y)],
\end{equation}
where the notation $tr$ includes trace over the Dirac, color and flavor indices and ${\cal{G}}^{-1}_{0}(x,y)$ denotes the inverse propagator with the self-energy $\Sigma(x,y)=\Lambda^{\theta} B^{\theta}_{0}(x,y)$ at the finite chemical potential $\mu$. Employing the stationary condition Eq.(7), and reversing the Fierz transformation, we have
\begin{equation}
\Sigma[\mu](y_1,y_2)=\frac{4}{3}g^2_{s} D(y_1,y_2)\gamma_{\nu}
{\cal{G}}_{0}[\mu](y_1,y_2)\gamma_{\nu}.
\end{equation}

It should be noted that both $B^{\theta}_{0}(x,y)$ and ${\cal{G}}^{-1}_{0}(x,y)$ are dependent on the chemical potential $\mu$. If the chemical potential $\mu$ is switched off, ${\cal{G}}_{0}[\mu]$ goes into the dressed vacuum quark propagator $G\equiv{\cal{G}}_{0}[\mu=0]$, which has the decomposition
\begin{equation}
G^{-1}(p)=i\gamma\cdot p+\Sigma (p)=i\gamma\cdot p A(p^2)+B(p^2)
\end{equation}
with
\begin{equation}
\Sigma(p)=\int d^4 x e^{i p\cdot x}[\Lambda^{\theta} B^{\theta}_{0}(x)]=\frac{4}{3}\int \frac{d^4 q}{(2\pi)^4} g^2_{s}D(p-q)\gamma_{\nu}G(q)\gamma_{\nu}
\end{equation}
where the self energy functions $A(p^2)$ and $B(p^2)$ are
determined by the rainbow Dyson-Schwinger equation(DSE)
\[
[A(p^2)-1]p^2=\frac{8}{3}\int \frac{d^{4}q}{(2\pi)^4}g^2 D(p-q)
\frac{A(q^2)p\cdot q}{q^2A^2(q^2)+B^2(q^2)},
\]
\begin{equation}
B(p^2)=\frac{16}{3}\int \frac{d^{4}q}{(2\pi)^4}g^2 D(p-q)
\frac{B(q^2)}{q^2A^2(q^2)+B^2(q^2)}.
\end{equation}

Here we want to stress that the $B(p^2)$ in Eq.(11) has two qualitatively distinct solutions. The ``Nambu-Goldstone'' solution, for which
\begin{equation}
B(p^2)\neq 0,
\end{equation}
describes a phase in which: 1) chiral symmetry is dynamically broken. Because one has a nonzero quark mass function; and 2) the dressed quarks are confined, because the propagator described by these functions does not have a Lehmann representation. The alternative ``Wigner'' solution, for which
\begin{equation}
B(p^2)\equiv 0,
\end{equation}
describes a phase in which chiral symmetry is not broken and the dressed-quarks are not confined. In ``Wigner'' phase, the Dyson-Schwinger equation(11) reduces to:
\[
[A'(p^2)-1]p^2=\frac{8}{3}\int \frac{d^{4}q}{(2\pi)^4}g^2 D(p-q)
\frac{p\cdot q}{q^2A'(q^2)},
\]
where $A'(p^2)$ denotes the dressed quark vector self energy function in ``Wigner'' phase. Therefore, the dressed quark propagator in ``Wigner'' phase can be written as $G^{(W)}(q)=\frac{-i\gamma\cdot q}{A'(q^2)q^2}$.

In order to get the numerical solution of $A(p^2)$, $B(p^2)$ and $A'(p^2)$, one often use model forms for gluon two-point function as input in Eq.(11). Here we investigate two different two parameter models for gluon propagator;
\begin{equation}
g^2D^{(1)}(q^2)=g^2D^{(1)}_{IR}(q^2)+g^2D_{UV}(q^2)=3\pi^{2}\frac{\chi^2}{\Delta^2}e^{-\frac{q^2}{\Delta}}+\frac{4\pi^2 d}{q^2ln\left(\frac{q^2}{\Lambda^2_{QCD}}+e\right)},
\end{equation}
and
\begin{equation}
g^2D^{(2)}(q^2)=g^2D^{(2)}_{IR}(q^2)+g^2D_{UV}(q^2)=4\pi^2 d\frac{\chi^2}{q^4+\Delta}+\frac{4\pi^2 d}{q^2ln\left(\frac{q^2}{\Lambda^2_{QCD}}+e\right)}.
\end{equation}

The term $D_{IR}(q^2)$, which dominates for small $q^2$, simulates the infrared enhancement and confinement. The dressed-gluon propagator is strongly enhanced, which leads via the QCD gap equation(11) to an infrared enhancement of the light quark mass function. These modification are intimately related to the confinement and dynamical chiral symmetry breaking[7]. The other term $D_{UV}(q^2)$, which dominates for large $q^2$, is an asymptotic ultraviolet(UV) tail match the known one-loop renormalization group result with $d=[12/(33-2N_{f})]=12/27$, $\Lambda_{QCD}=200~MeV$. The model parameters $\chi$ and $\Delta$ are adjusted to reproduce the weak decay constant in the chiral limit $f_{\pi}=87~MeV$. The forms of $g^2D(q^2)$ have been used in Ref.[8] and it has been shown that with these values a satisfactory description of all low energy chiral observables can be achieved(more detail can be seen in Refs.[8] and [9]).

Let us now study the chemical potential dependence of the dressed quark propagator. To this, one can numerically study Eq.(8). From Lorentz structure, the most general form for the ${\cal{G}}^{-1}(p,\mu)$(in the chiral limit) which fulfills Eq.(8), reads
\begin{eqnarray}
{\cal{G}}^{-1}(p,\mu)&=&i\gamma\cdot p~a(p^2,u^2,[u\cdot p]^2)+b~(p^2,u^2,[u\cdot p]^2)+i\gamma\cdot u p\cdot u~c(p^2,u^2,[u\cdot p]^2)\\
&&+i\gamma\cdot p p\cdot u~d(p^2,u^2,[u\cdot p]^2)-\gamma\cdot u~e(p^2,u^2,[u\cdot p]^2)+i p\cdot u~f(p^2,u^2,[u\cdot p]^2),\nonumber
\end{eqnarray}
where $u_{\mu}=(\vec{0},\mu)$. However, the model gluon propagator(14,15), as pointed out above,  has no explicit $\mu$-dependence, which can arise through quark vacuum polarisation insertions. As such it may be inadequate at large value of $\mu$, particularly near any critical chemical potential. Therefore, it is comparatively safe to study the low chemical potential dependence of the dressed quark propagator by means of Eq.(8) with the model gluon propagator(14,15). Here we restrict ourselves to study the first-order dependence of ${\cal{G}}^{-1}_{0}[\mu]$ upon $\mu$. Namely, we can expands ${\cal{G}}^{-1}_{0}[\mu]$ in powers of $\mu$ as follows
\begin{equation}
{\cal{G}}^{-1}_{0}[\mu]=\left.{\cal{G}}^{-1}_{0}[\mu]\right|_{\mu=0}+\left. \frac{\delta {\cal{G}}^{-1}_{0}[\mu]}{\delta {\mu}} \right|_{\mu=0}\mu+
{\cal{O}}(\mu^2)=G^{-1}+\mu\Gamma_{4}+{\cal{O}}(\mu^2),
\end{equation}
which leads to the formal expansion
\begin{equation}
{\cal{G}}_{0}[\mu]=G-G\mu\Gamma_{4}G+\cdot\cdot\cdot,
\end{equation}
with $\Gamma_{4}$
\begin{equation}
\Gamma_{4}(y_1,y_2)=\left. \frac{\delta {\cal{G}}^{-1}_{0}[\mu](y_1,y_2)}
{\delta \mu} \right|_{\mu=0}.
\end{equation}
In coordinate space the dressed vertex $\Gamma_{4}(x,y)$ is given as the 
derivative of the inverse quark propagator ${\cal{G}}^{-1}_{0}[\mu]$ with respect to the chemical potential $\mu$. 

Taking the derivative in Eq.(5) and putting it into Eq.(19), we have
\begin{equation}
\Gamma_{4}(y_1,y_2)=-\gamma_{4}\delta(y_1-y_2)+
\left. \frac{\delta\Sigma[\mu](y_1,y_2)}{\delta{\mu}} \right|_{\mu=0}.
\end{equation}

Substituting Eq.(8) and (18) into Eq.(20), we have the inhomogeneous ladder Bethe-Salpeter equation(BSE) for $\Gamma_{4}$ vertex, which reads
\begin{eqnarray}
\Gamma_{4}(y_1,y_2)&&=-\gamma_{4}\delta(y_1-y_2)\nonumber\\
&&-\frac{4}{3}g^2_{s}D(y_1-y_2)\int du_{1}du_{2}\gamma_{\nu}G(y_1,u_1)
\Gamma_{4}(u_1,u_2)G(u_2,y_2)\gamma_{\nu}
\end{eqnarray}

Fourier transform of Eq.(21) leads then to the momentum space form of $\Gamma_{4}$

\begin{equation}
\Gamma_{4}(P,0)
=-\gamma_{4}
-\frac{4}{3}\int\frac{d^4 K}{(2\pi)^4}g^2_{s}D(P-K)\gamma_{\nu}G(K)\Gamma_{4}
(K,0)G(K)\gamma_{\nu}.
\end{equation}

Eqs.(9) and (22) yields 
\begin{equation}
\Gamma_{4}(P,0)=i\frac{\partial G^{-1}(P)}{\partial P_{4}};
\end{equation}
i.e., the vector ``Ward identity'' is satisfied[10,11]. This means that it is the nonperturbative dressed effect which modifies the bare vertex $\gamma_{4}$ to the nonperturbative dressed vertex $\Gamma_{4}(P,0)$ at the level of linear response approximation.

By means of ``Ward identity''(23) and Eq.(17), we have the chemical potential dependence of the dressed quark propagator in ``Nambu-Goldstone'' and ``Wigner''phase separately( here we only consider the first order dependence of ${\cal{G}}^{-1}_{0}[\mu]$ upon $\mu$ at the mean field level);
\begin{eqnarray}
{{\cal{G}}_{0}^{(NG)}}^{-1}[\mu]&=& i\vec{\gamma}\cdot\vec{p}\left[A(P^2)+2i\mu P_4\frac{\partial A(P^2)}{\partial P^2}\right]+B(P^2)+2i\mu P_4\frac{\partial B(P^2)}{\partial P^2}\nonumber\\
&&+i\gamma_4\left[P_4A(P^2)+2i\mu P^2_4\frac{\partial A(P^2)}{\partial P^2}+
i\mu A(P^2)\right]+{\cal{O}}(\mu^2),
\end{eqnarray}
\begin{eqnarray}
{{\cal{G}}_{0}^{(W)}}^{-1}[\mu]&=&i\vec{\gamma}\cdot\vec{p}\left[A'(P^2)+2i\mu P_4\frac{\partial A'(P^2)}{\partial P^2}\right]\nonumber\\
&&+i\gamma_4\left[P_4A'(P^2)+2i\mu P^2_4\frac{\partial A'(P^2)}{\partial P^2}+
i\mu A'(P^2)\right]+{\cal{O}}(\mu^2).
\end{eqnarray}
Note that the functions $A(P^2)$, $B(P^2)$ and $A'(P^2)$ here, as pointed out above, are known by numerically solving the corresponding DSE.

As it is shown in Eqs.(24) and (25), for $\mu\not= 0$ the dressed-quark self energies in general acquire an imaginary part driven by the chemical potential $\mu$[12-15]. Just as pointed out in Ref.[6], this effect is not observed in the study of the Nambu-Jona-Lasinio model and the instanton-induced four fermion interaction model in which the interaction is energy-independent; i.e., instantaneous. In addition, it should be noted that the above approach for getting the nonperturbative $\Gamma_4$ vertex has been proven to be very useful for the studies of nonperturbative vector and axial vector vertex[10-11,16]. This approach employs a consistent treatment of the dressed quark propagator $G$ and the dressed vertex $\Gamma_4$, which are both determined from the effective quark-quark interaction by the rainbow DSE for $G$ and the inhomogeneous ladder BSE for $\Gamma_{4}$.

With these two ``phase'' characterized by qualitatively different momentum-dependent quark propagator(24,25), the GCM can be used to explore chiral symmetry restoration and phase transition between ``Wigner'' and ``Nambu-Goldstone'' phase.

To explore the possibility of a phase transition one must consider the relative stability of the confined and deconfined phase by computing the $\mu$-dependence of vacuum pressure difference(or ``bag constant''[3]). It is equivalent to calculating the difference between the tree-level auxiliary-field effective action[17] evaluated with the ``Wigner'' solution characterised by $B(p^2)\equiv 0$, and the ``Nambu-Goldstone'' solution characterised By $B(p^2)\neq 0$[12]:
\begin{eqnarray}
&&{\cal{B}}(\mu)\equiv P[{\cal{G}}_{0}^{(NG)}]-P[{\cal{G}}_{0}^{(W)}]\nonumber\\
&&=4N_{C}\int\frac{d^4p}{(2\pi)^4}\left\{ln\left |\frac{A^2\vec{p}^2+C^2+B^2}{A'^{2}\vec{p}^2+C'^2}\right |+Re\left[\vec{p}^2(\sigma_{A}-\sigma'_{A})+(p_4+i\mu)(\sigma_{C}-\sigma'_{C})\right]\right\}\nonumber\\
&&=\frac{3}{2\pi^3}\int sds\int^{\pi}_{0} sin^2\chi d\chi\left\{ln\left |\frac{sA^2sin^2\chi+C^2+B^2}{sA'^2sin^2\chi+C'^2}\right |\right.\\
&&\left.+Re\left[s~sin^2\chi\left(\sigma_{A}-\sigma'_{A}\right)+\left(s^{1/2}cos\chi+i\mu\right)\left(\sigma_{C}-\sigma'_{C}\right)\right]\right\},\nonumber
\end{eqnarray}
with\[ A=[A(s)+2i\mu s^{1/2}cos\chi\frac{\partial A(s)}{\partial s}],~C=s^{1/2}cos\chi[A(s)+2i\mu s^{1/2}cos\chi\frac{\partial A(s)}{\partial s}]+i\mu A(s),\]
\[A'=[A'(s)+2i\mu s^{1/2}cos\chi\frac{\partial A'(s)}{\partial s}],~C'=s^{1/2}cos\chi[A'(s)+2i\mu s^{1/2}cos\chi\frac{\partial A'(s)}{\partial s}]+i\mu A'(s),\]
\[B=B(s)+2i\mu s^{1/2}cos\chi\frac{\partial B(s)}{\partial s},~\sigma_{A}=\frac{A}{s~sin^2\chi A^2+C^2+B^2},~\sigma'_{A}=\frac{A'}{s~sin^2\chi A'^2+C'^2},\]
\[\sigma_{C}=\frac{C}{s~sin^2\chi A^2+C^2+B^2},~\sigma'_{C}=\frac{C'}{s~sin^2\chi A'^2+C'^2}.\]

${\cal{B}}(\mu)>0$ indicates the stability of the confined(Nambu-Goldstone) phase and hence the phase boundary is specified by ${\cal{B}}(\mu)=0$. ${\cal{B}}(\mu)/{\cal{B}}(0)$ is plotted in Figs.1-2. The scale is calculated to be ${\cal{B}}(0)=0.131 \sim 0.152~(GeV)^4$(see 
Table.I), which can be compared with the value $(0.145~GeV)^4$ commonly used in bag-like models of hadron[18]. It is positive when the Nambu-Goldstone phase is dynamically favoured; i.e., has the highest pressure and become negative when the Wigner pressure become larger. The critical chemical potential is the zero of ${\cal{B}}(\mu)$: i.e., $\mu_{c}=270 \sim 375~ MeV$. This abrupt switch from the Nambu-Goldstone to the Wigner phase signals a first order transition. It should be noted that our numerical results is only valid for small values of chemical potential. If it could be qualitatively extrapolated to the reasonably large $\mu_{c}$, then our conclusion about the first order phase transition is correct.

\begin{center}

\epsfig{file=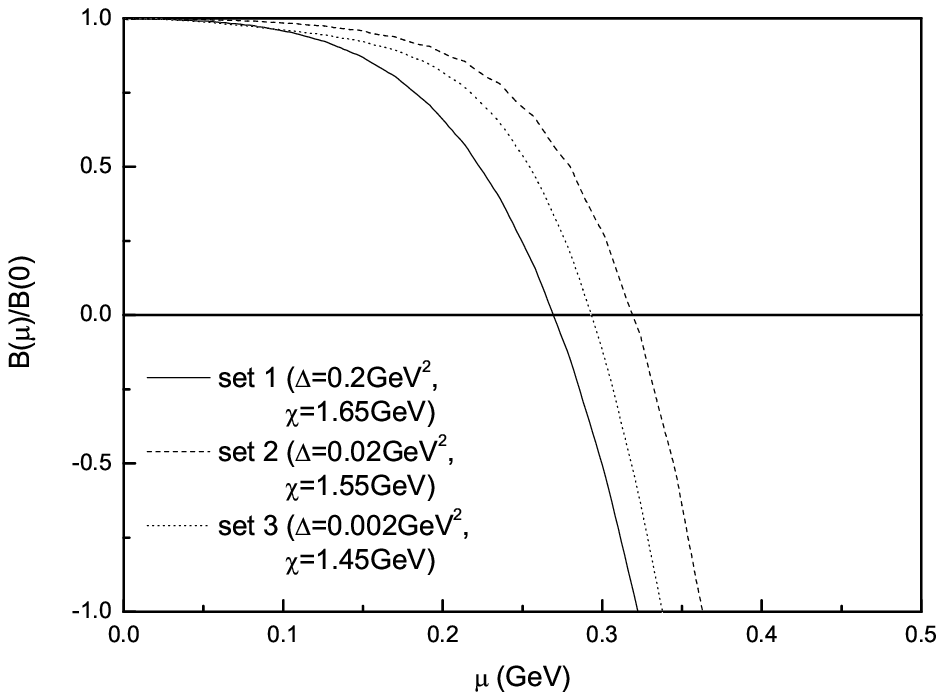, width=10cm}

\vspace{-0.8cm}

Fig.1. Ratio $B(\mu)/B(0)$ for gluon propagator Eq.(14).

\epsfig{file=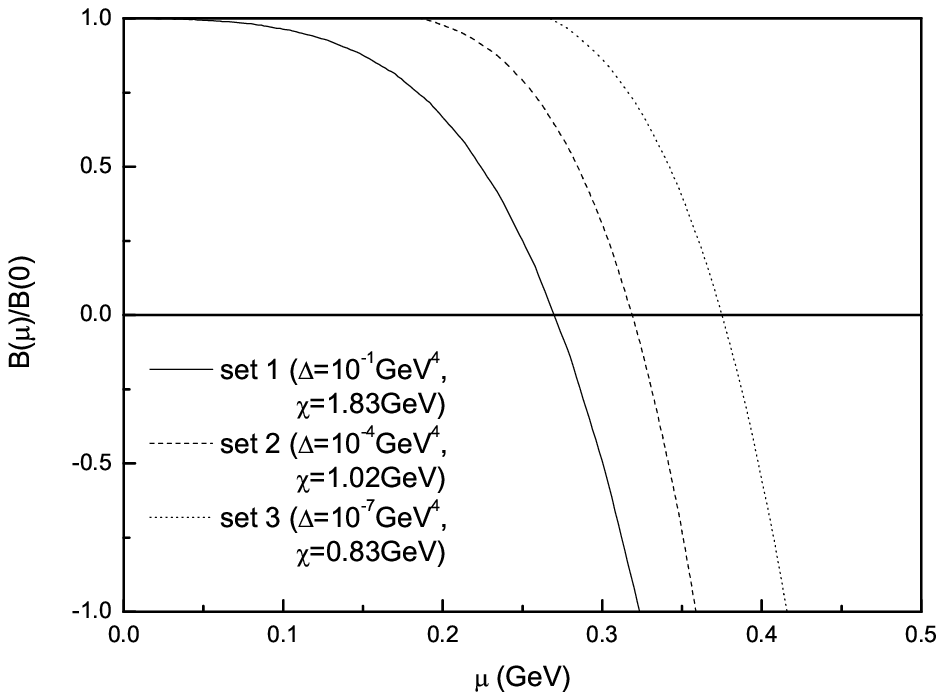, width=10cm}

\vspace{-0.8cm}

Fig.2. Ratio $B(\mu)/B(0)$ for gluon propagator Eq.(15).

\end{center}
\begin{center}
\begin{tabular}{ccc|ccc} 
\multicolumn{6}{c}{Table. I. The bag constant for two different models gluon propagator}\\ \hline\hline   
\multicolumn{3}{c|}{$g^2D^{(1)}_{IR}(q^2)$=$3\pi^2\frac{\chi^2}{\Delta^2}e^{\frac{q^2}{\Delta}}$}&
\multicolumn{3}{c}{$g^2D^{(2)}_{IR}(q^2)$=$4\pi^2 d\frac{\chi^2}{q^4+\Delta}$}\\ \hline
$\Delta[GeV^2]$ &~~~ $\chi[GeV]$ ~~~ & ${\cal{B}}^{\frac{1}{4}}(0)[GeV]$ & $\Delta[GeV^4]$ & $\chi[GeV]$ & ${\cal{B}}^{\frac{1}{4}}(0)[GeV]$ \\ \hline
0.200  &1.65   & 0.136    & $10^{-1}$ &1.83 &0.135  \\
0.020  &1.55   & 0.139    & $10^{-4}$ &1.02 &0.136  \\
0.002  &1.45   & 0.131    & $10^{-7}$ &0.83  &0.152 \\ \hline\hline
\end{tabular}
\end{center}

The chiral quark condensate is proportional to the matrix trace of the chiral-limit dressed quark propagator. Using Eq.(24), we obtain the following expression valid in the domain of confinement and DCSB:
\begin{equation}
\langle\tilde{0}|:\bar{q}q:|\tilde{0}\rangle_{\mu}=-tr_{DC}
\left\{Re~{\cal{G}}^{(NG)}_{0}[\mu]\right\}.
\end{equation}
The calculated ratio $\langle\tilde{0}|:\bar{q}q:|\tilde{0}\rangle_{\mu}/\langle\tilde{0}|:\bar{q}q:|\tilde{0}\rangle$ can be seen from the Figs.3-4. In Figs.3-4, we see that $\langle\tilde{0}|:\bar{q}q:|\tilde{0}\rangle_{\mu}$ decreases with increasing $\mu$, up to a critical value of $\mu_{c}$, when it drops continuously to zero(It should be noted that although our numerical result is only valid for low values of chemical potential, we may still expect that it qualitatively gives the reasonable large $\mu$ tendency of the two-quark condensate). The change of $\langle\tilde{0}|:\bar{q}q:|\tilde{0}\rangle_{\mu}$ with increasing $\mu$ in our calculation is consistent with results in Ref.[19-20]. However, this result is qualitatively different from that in Refs.[13,14]. In Refs.[13,14], the chiral order parameter increases with increasing chemical potential up to $\mu_c$. At $\mu_c$, it drops abruptly to zero. We want to stress that the conclusion of Refs.[12-14] is based on the fact that those authors adopt the following general form for the solution of Eq.(8):  
\begin{equation}
{\cal{G}}^{-1}(\tilde{p})=i\gamma\cdot p A(\tilde{p})+i\gamma_4(p_4+i\mu)C(\tilde{p})+B(\tilde{p}),
\end{equation}
where $\tilde{p}\equiv (\vec{p}, p_4+i\mu)$. Although not explicitly indicated, the solutions($A(\tilde{p}), B(\tilde{p})$ and $C(\tilde{p})$) are functions only of $|\vec{p}|^2$ and $(p_4+i\mu)^2$[14]. Comparing Eq.(24) with Eq.(28), it is easily seen that our solution(24) can not be included in the general form(28) given by Refs.[12-14].

\begin{center}
\epsfig{file=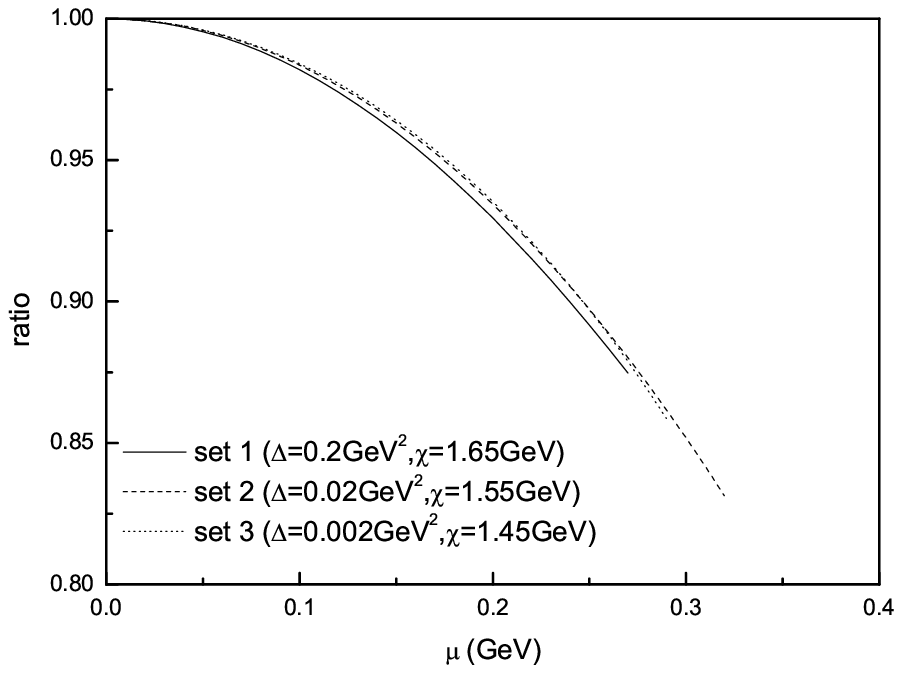, width=11cm,}

\vspace{-0.5cm}

Fig.3. Ratio $\langle\tilde{0}|:\bar{q}q:|\tilde{0}\rangle_{\mu}/\langle\tilde{0}|:\bar{q}q:|\tilde{0}\rangle$ for gluon propagator Eq.(12).

\epsfig{file=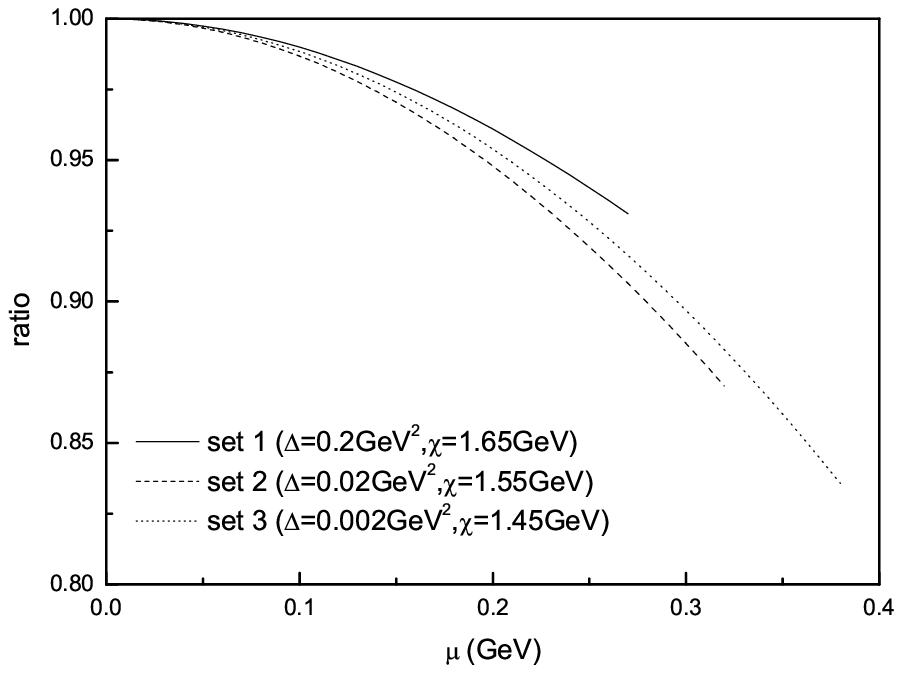, width=11cm,}

\vspace{-0.5cm}

Fig.4. Ratio $\langle\tilde{0}|:\bar{q}q:|\tilde{0}\rangle_{\mu}/\langle\tilde{0}|:\bar{q}q:|\tilde{0}\rangle$ for gluon propagator Eq.(13).

\end{center}

To summarize: in the present paper, we provide a general recipe to calculate the low chemical potential dependence of the dressed quark propagator at the mean field level in the framework of GCM. From this the $\mu$ dependence of the quark condensate and bag constant are evaluated. It is found that the chiral order parameter decreases with increasing $\mu$, which is qualitatively different from that in Refs.[13,14]. The reason for this discrepancy is discussed.

\vspace*{0.2 cm}
\noindent{\large \bf Acknowledgments}

This work was supported in part by the National Natural Science Foundation
of China under Grant Nos 19975062, 10175033 and 10135030.

\vspace*{0.2 cm}
\noindent{\large \bf References}

\begin{description}
\item{[1]} T. D. Lee, Nucl. Phys.
{\bf A590}, 11c (1995).
\item{[2]} M. P. Lombardo, J. B. Kogut, D. K. Sinclair, Phys. Rev.
{\bf D54},
 2303 (1996).
\item{[3]} R. T. Cahill and C. D. Roberts, Phys. Rev. {\bf D32}, 2419 (1985).
\item{[4]} P. C. Tandy, Prog. Part. Nucl. Phys. 39, 117 (1997); R. T. Cahill and S. M. Gunner, Fiz. {\bf B7}, 17 (1998), and references therein.
\item{[5]} C. D. Roberts and A. G. Williams, Prog. Part. Nucl. Phys. {\bf 33}, 477 (1994), and references therein.
\item{[6]} C. D. Roberts and S. M. Schmidt, Prog. Part. Nucl. Phys. {\bf 45}, 
1 (2000), and references therein.
\item{[7]} A.G. Williams, G. Krein, C. D. Roberts, Ann Phys. {\bf 210}, 464 (1991); H. J. Munezek, P. Jain, Phys. Rev. {\bf D28}, 438 (1992).
\item{[8]} M. R. Frank and T. Meissner, Phys. Rev {\bf C53}, 2410 (1996).
\item{[9]} T. Meissner, Phys. Lett. {\bf B405}, 8 (1997).
\item{[10]} M. R. Frank, Phys. Rev. {\bf C51}, 987 (1995).
\item{[11]} T. Meissner and L. S. Kisslinger,  Phys. Rev. {\bf C59}, 986 (1999).
\item{[12]} D. Blaschke, C. D. Roberts, S. Schmidt, Phys. Lett. {\bf B425}, 232 (1998).
\item{[13]} P. Maris, C. D. Roberts, and S. Schmidt, Phys Rev. {\bf C57}, R2871 (1998).
\item{[14]} A. Bender, G. I. Poulis, C. D. Roberts, S. Schmidt, A. W. thomas, Phys. Lett. {\bf B431}, 263 (1998).
\item{[15]} Yu-xin Liu, Dong-feng Gao, Hua Guo, Nucl. Phys. {\bf A695}, 353 (2001).
\item{[16]} Hong-shi Zong, Xiang-song Chen, Fan Wang, Chao-hsi Chang and En-guang Zhao, Phys. Rev. {\bf C66}, 015201 (2002).
\item{[17]} R. W. Haymaker, Riv. Nuovo Cim. 14 (1991) series 3. no. 8.
\item{[18]} R. T. Cahill, Aust. J. Phys. {\bf 42}, 171 (1989).
\item{[19]} W. Weise, Nucl. Phys. {\bf A610}, 35c (1996).
\item{[20]} S. P. Klevansky, ReV. Mod. Phys. {\bf 64}, 649 (1992), and references therein.
\end{description}

\end{document}